\documentclass[nofootinbib,showpacs,aps]{revtex4}
\usepackage{graphicx}
\begin{document}
\def\be{\begin{equation}}
\def\ee{\end{equation}}
\def\bea{\begin{eqnarray}}
\def\eea{\end{eqnarray}}
\title{Relativisticlike structure of classical thermodynamics}
\author{Hernando Quevedo$^{1,2}$, Alberto S\'anchez$^3$ and  Alejandro V\'azquez$^1$}
\email{quevedo@nucleares.unam.mx,asanchez@nucleares.unam.mx,alec_vf@nucleares.unam.mx}
\affiliation{$^1$Instituto de Ciencias Nucleares, 
Universidad Nacional Aut\'onoma de M\'exico, 
 AP 70543, M\'exico, DF 04510, Mexico}
\affiliation{$^2$ 
Dipartimento di Fisica and ICRA, Universit\`a di Roma ``La Sapienza", Piazzale Aldo Moro 5, I-00185 Roma, Italy}
\affiliation{$^3$ Departamento de Posgrado, CIIDET, AP 752, Quer\'etaro, QRO 76000, Mexico}

\begin{abstract}

We analyze in the context of geometrothermodynamics a Legendre invariant  metric structure in the equilibrium space of an ideal gas. 
We introduce the concept of thermodynamic geodesic as a succession of points, each corresponding to a state of equilibrium, 
so that the resulting curve represents a quasi-static process. 
A rigorous geometric structure is derived in which the thermodynamic geodesics at a given point split the equilibrium space into two disconnected regions 
separated by adiabatic geodesics. This resembles the causal structure of special relativity, which we use to introduce the concept 
of adiabatic cone for thermodynamic systems. This result might be interpreted as an alternative indication of the 
inter-relationship between relativistic physics and classical thermodynamics.

\end{abstract}
\pacs{04.20. -q, 02.40.-k, 05.70.-a}

\maketitle

\section{Introduction}
\label{sec:int}

The not yet understood similarity between the laws of black hole dynamics and classical thermodynamics (for a recent review, see \cite{wald01}) seems to indicate that there exists a deep relationship between gravitational dynamics and thermodynamics. This relationship was also established in flat spacetime in the sense that thermodynamic variables like temperature can be ascribed to any null surface that acts as a horizon for observers moving along specific accelerated trajectories in flat spacetime (Unruh effect) \cite{unruh76}. In all these examples, the null surface that can be interpreted as a horizon seems to play an essential role. Nevertheless, other examples are known in which no horizon exists, for instance in a spherically symmetric perfect fluid in equilibrium, but certain similarity continues to hold in the sense that Einstein's equations are a consequence of demanding that the total entropy of the matter be an extremal \cite{open01}. 

The inter-relationship between gravity and thermodynamics was further strengthened by the discovery that Einstein's equations can be interpreted as a first law of thermodynamics \cite{jac95}. Additional studies of the dynamics of gravity have shown that the validity of this relationship is wider than originally suspected and it holds also in other gravity theories (for a review, see \cite{pad13}). These results suggest that gravity is not a fundamental interaction, but is an emergent phenomenon with its field equations having the same status as the equations of fluid dynamics. If this turns out to be true, far-reaching consequences are to be expected, in particular, regarding the quantization of gravity.  

In all the above studies, one stars from a gravitational system and tries to attribute thermodynamic variables in a consistent manner in order to find the analogy between the dynamics of the system  and classical thermodynamics. In this work, we explore the connection between relativity and thermodynamics from a different perspective, namely, we start from a thermodynamic system and try to find the analogy between its ``dynamics" and relativity. Indeed, we will show that in the space of equilibrium states of the ideal gas, quasi-static processes can be represented as geodesics which induce a causality structure, resembling that of special relativity. To this end, it is necessary to interpret the equilibrium space as a differential manifold, i.e., we need a geometric approach to thermodynamics.

The idea of using differential geometry in thermodynamics is due to Gibbs \cite{gibbs}
who realized that the first law of thermodynamics can be represented in terms of
differential forms. Caratheodory \cite{car} interpreted 
the laws of thermodynamics in an axiomatic way and in terms of Pfaffian forms. 
Later on, Hermann \cite{her} introduced the concept of thermodynamic phase space ${\cal T}$ 
where the thermodynamic variables play the role of coordinates. In turn, the phase space contains, in principle, an infinite number of 
  subspaces ${\cal E} \subset {\cal T}$ called equilibrium spaces, each of which is determined by a particular fundamental equation. 
Recall that in classical thermodynamics, the properties of any thermodynamic system are completely described by the corresponding fundamental equation \cite{callen}. 
The points of each equilibrium space are interpreted as equilibrium states at which the system can possibly exist. 

In addition, in the phase space
one can introduce a contact structure \cite{her} which is invariant with respect to Legendre transformations.
The connection between the structure of the phase space
and the equilibrium space ${\cal E}\subset {\cal T}$ was recently incorporated  in the formalism 
of geometrothermodynamics (GTD) in an invariant manner  \cite{quev07}.
In classical thermodynamics, different thermodynamic potentials are related by means of Legendre
transformations \cite{callen}. Consequently, the Legendre invariance of GTD guarantees that the properties 
of a thermodynamic system do not depend on the thermodynamic potential used for its description.
GTD delivers for any thermodynamic system a Legendre 
invariant metric which describes the geometry of ${\cal E}$. 
 In this manner, the equilibrium space becomes a Riemannian manifold whose geometric properties should be related to the properties
of the corresponding thermodynamic system.

In GTD, we use some intuitive geometric concepts that were first used in general relativity. 
For instance, the curvature tensor is expected to be
a measure of the interaction between the components of the thermodynamic system. In the case of an ideal gas,
 where the interaction between particles can be neglected, we say that
the thermodynamic interaction is zero and, consequently, the curvature must vanish. 
We will present a metric for the equilibrium space whose curvature satisfies this condition. 
Moreover, the curvature tensor is not zero for systems with intrinsic thermodynamic interaction. For this reason  
the curvature of the equilibrium space is called {\it thermodynamic curvature}. In general relativity,
curvature singularities indicate the break down of the theory. We use the same idea in GTD and associate 
the singularities of the thermodynamic curvature with the occurrence of phase transitions at which the equilibrium 
approach of classical thermodynamics is no more valid. 

In this work, we introduce the concept of {\it thermodynamic geodesic} as a solution of the geodesic equations along which the laws of thermodynamics are fulfilled. 
We show that thermodynamic geodesics describe quasi-static processes and that the affine parameter can be used as
a ``time" parameter. Moreover, thermodynamic geodesics have a definite direction that can be interpreted as the ``arrow of time". 
In the case of an ideal gas, we show that the thermodynamic geodesics split the equilibrium space into two non-connected regions that resemble the causality structure in Minkowski spacetime.

This paper is organized as follows. In Sec. \ref{sec:gtdbh}, we present the fundamentals of GTD in the case 
of systems with two thermodynamic degrees of freedom. In particular, we present a Legendre invariant metric for
the phase space from which it is possible to derive an invariant geometry for the space of equilibrium. 
In Sec. \ref{sec:ig}, we focus on the geometry of the ideal gas and show that the
corresponding thermodynamic curvature vanishes as a result of the lack of thermodynamic interaction.
Furthermore, in Secs. \ref{sec:geoig} and \ref{sec:cone}, we perform a detailed analysis of the thermodynamic geodesics 
of the ideal gas. The equilibrium space presents a very rich and unexpected structure that resembles the 
causality structure of spacetime in relativistic physics. In particular, we use the entropy as an affine parameter along thermodynamic geodesics 
to introduce at each equilibrium state an ``adiabatic cone" that determines the entropic past and future of that state.   
 Finally, Section \ref{sec:con} is devoted
to discussions of our results and suggestions for further research.


\section{Geometrothermodynamics of simple systems}
\label{sec:gtdbh}

A rigorous formulation of GTD implies the introduction of geometrical concepts like
contact manifolds, tangent and cotangent contact manifolds, smooth maps, pullbacks, etc. \cite{quev07,qq11,blnq14}. 
The main purpose of this section is to present GTD in a less rigorous manner, using only the essential tools of differential geometry and emphasizing the 
intuitive physical and geometric aspects.

The simplest thermodynamic systems are those with only two thermodynamic degrees 
of freedom. To describe this kind of systems, it is necessary to specify two 
extensive variables, say entropy $S$ and volume $V$, two intensive variables, 
say temperature $T$ and pressure $P$, and a thermodynamic potential, say the
internal energy $U$. In GTD \cite{quev07}, the thermodynamic variables are used as 
coordinates for the construction of the thermodynamic phase space ${\cal T}$,
which in this case is 5-dimensional. We assume that ${\cal T}$ is a well-behaved
differential manifold so that it allows us to introduce additional 
geometric structures. In particular, we introduce the so-called fundamental 
Gibbs form
\be
\label{gibbs}
\Theta = dU - T d S + P d V \ ,
\ee
which, as we will see below, contains the information about the first law of 
thermodynamics. In addition, we introduce a metric structure $G$ that, in
general, can depend on all the coordinates of ${\cal T}$, i.e. $G=G(U,S,V,T,P)$.

The triplet $({\cal T},\Theta, G)$ is called a Riemannian contact manifold 
and represents an auxiliary structure which is necessary to implement 
in a consistent manner the properties of classical thermodynamics. In particular,
it is known that the thermodynamic properties of ordinary thermodynamic systems do not depend on 
the thermodynamic potential used for its description \cite{callen}. 
Since different thermodynamic
potentials are related by means of Legendre transformations, we demand that 
the geometric structure of the phase space ${\cal T}$ be Legendre invariant. 
In the case of a system with two degrees of freedom, 
a Legendre transformation is defined in ${\cal T}$ 
as the change of coordinates $(U,S,V,T,P)\longrightarrow 
(\tilde U,\tilde S,\tilde V,\tilde T,\tilde P)$ with the following 
possibilities \cite{arnold}:
\bea
\label{lt1}
\tilde U_1 = U - TS \ ,\quad S = -\tilde T\ ,\quad T =\tilde S \ ,\quad V = \tilde V
\ ,\quad P =\tilde P \ ,\\
\label{lt2}
\tilde U_2 = U + PV \ ,\quad S = \tilde S\ ,\quad T =\tilde T \ ,\quad V = \tilde P
\ ,\quad - P =\tilde V \ ,\\
\label{lt3}
\tilde U_3 = U - TS + PV \ ,\quad S = -\tilde T\ ,\quad T =\tilde S \ ,\quad V = \tilde P
\ ,\quad - P =\tilde V \ .
\eea
Usually, $\tilde U_1=F$ is called the Helmholtz free energy, $\tilde U_2=H$ is 
the enthalpy, and $\tilde U_3=G$ is the Gibbs potential. If we denote by $\tilde \Theta_i$, 
$i=1,2,3$, the result of applying any of the particular Legendre transformations 
(\ref{lt1})-(\ref{lt3}) to the fundamental Gibbs form, then it is easy to see that
$\tilde \Theta_i = d\tilde U_i - \tilde T d \tilde S +\tilde P d\tilde V$, showing that
in fact $\Theta$ is a Legendre invariant geometric object. 

In GTD, we also demand that the metric $G$ be Legendre invariant. It is possible to write
down and solve the algebraic conditions that an arbitrary metric $G$ must satisfy 
in order to be Legendre invariant \cite{quev07}. A particular solution was found in 
\cite{vqs08} that can be written as
\be
\label{ginv3}
G=\left(dU - T d S + P d V\right)^2 + (ST)^{2k+1} dS dT + (VP)^{2k+1} dV dP \ ,
\ee
where $k$ is an integer, positive or negative. 
As in the case of the
fundamental Gibbs form, if we denote by $\tilde G_i$
the metric resulting from the application of the Legendre transformations (\ref{lt1})-(\ref{lt3})
to the metric (\ref{ginv3}), the Legendre invariance of $G$ becomes clear. 
In fact, the functional dependence of $\tilde G_i$  coincides with (\ref{ginv3}) with 
$U$ replaced by $\tilde U_i$, $S$ by $\tilde S$, and so on. In this manner, we see that 
the particular triplet $({\cal T}, \Theta, G)$, with $\Theta$ and $G$ given as in Eqs.(\ref{gibbs}) and
(\ref{ginv3}), respectively, is invariant with respect to all possible Legendre transformations
in the case of systems with two thermodynamic degrees of freedom\footnote{It should be noticed that the metric (\ref{ginv3}) is not unique. One still can multiply $G$ by an arbitrary Legendre invariant function $\Lambda$ 
which, however, can be associated with a change of representation that should not affect the geometric properties of ${\cal T}$ \cite{blnq14}. Moreover, all mixed terms 
of the form $(SV)^{2k+1} dS dV + (SP)^{2k+1} dS dP + $ etc. can be added to the right-hand side of Eq.(\ref{ginv3}), without changing its symmetry properties. These additional terms, however, do not satisfy the additional condition that the metric $G$ must generate a flat metric for the ideal gas \cite{gl14}.}.
This is an important property which 
guarantees that our further results are independent of the choice of thermodynamic potential.

The next important element of GTD is the space of equilibrium states ${\cal E}$ which, in quite
general terms, is the space where systems in thermodynamic equilibrium can exist and 
their properties can be 
investigated. This means that ${\cal E}$ is a 2-dimensional subspace of ${\cal T}$ which we define in the following 
manner. Let us choose the set of extensive variables $(S,V)$ as the coordinates of ${\cal E}$. Then, when evaluated 
on ${\cal E}$, the remaining coordinates of ${\cal T}$ must be functions of $S$ and $V$ only, i. e.
\be
\label{coor1}
U=U(S,V)\ ,\quad T=T(S,V)\ ,\quad P=P(S,V)\ .
\ee
The first of these equations is known as the fundamental equation in the energy representation. In fact, once 
$U(S,V)$ is given explicitly, one can derive all the equations of state and thermodynamic properties of the
corresponding thermodynamic system.  To guarantee the existence of the second and third equations of (\ref{coor1}), we
demand that the projection of the fundamental Gibbs form on ${\cal E}$ vanishes, i. e.,
\be
\label{firstlaw}
\Theta |_{\cal E} = 0 \Longleftrightarrow dU = T d S - P d V \ ,
\ee
a relationship that is immediately recognized as the first law of thermodynamics. Furthermore, since $U=U(S,V)$, the first
law of thermodynamics implies that
\be
\label{eqcond}
\frac{\partial U}{\partial S} = T\ ,\quad \frac{\partial U}{\partial V} = - P\ ,
\ee
so that $T$ and $P$ become functions of $S$ and $V$, as stated in (\ref{coor1}). In classical thermodynamics, the relationships
(\ref{eqcond})
represent the conditions for thermodynamic equilibrium. As for the metric $G$ of ${\cal T}$, we demand that its projection
on ${\cal E}$, by using (\ref{firstlaw}) and (\ref{eqcond}),   induces a metric $g$ on ${\cal E}$, i. e., $G|_{\cal E} = g = g(S,V)$.
In the particular case of the Legendre invariant metric (\ref{ginv3}), a straightforward calculation leads to
\bea
g  = && 
\left(S\frac{\partial U}{\partial S}\right)^{2k+1}
\frac{\partial^2 U}{\partial S ^2} d S^2
+ \left(V\frac{\partial U}{\partial V}\right)^{2k+1} 
\frac{\partial^2 U}{\partial V ^2} d V^2 \nonumber \\
\label{gdowninv1}
& & +   \left[ \left(S\frac{\partial U}{\partial S}\right)^{2k+1}
+\left(V\frac{\partial U}{\partial V}\right)^{2k+1}\right]
\frac{\partial^2 U}{\partial S \partial V} d S d V   \ .
\eea

If we specify a fundamental equation $U=U(S,V)$, the above metric is unique. 
The above description corresponds to the energy representation in classical thermodynamics. One of the advantages of GTD is that the formalism
allows us to handle different representations in an invariant way. For later purposes, we present here the entropy representation for a system
with two thermodynamic degrees of freedom. A simple rearrangement of Eq.(\ref{gibbs}) leads to the Gibbs form 
in the entropy representation
\be
\Theta_{_S} = dS - \frac{1}{T}dU - \frac{P}{T} dV \ ,
\ee
so that the coordinates of the phase space ${\cal T}$ are $(S,U,V,1/T,P/T)$, 
and the metric $G$ as given in (\ref{ginv3}) can be written as
\be
\label{gups}
G_{_S} = \left(dS -\frac{1}{T} d U - \frac{P}{T} dV\right)^2
+ \left(\frac{U}{T}\right)^{2k+1} dU d\left(\frac{1}{T}\right) 
+\left(\frac{VP}{T}\right)^{2k+1} dV d\left(\frac{P}{T}\right) \ .
\ee
Applying the corresponding change of coordinates (see Appendix), 
it is easy to show that the above geometric objects are invariant
with respect to Legendre transformations. Furthermore, for the equilibrium subspace ${\cal E}$ we choose the
extensive variables $U$ and $V$ so that the remaining coordinates become functions of $U$ and $V$ when projected 
on ${\cal E}$. In particular, the fundamental equation must now be given as $S=S(U,V)$. As before, we demand that
the projected Gibbs form and the metric satisfy the relationships
\be
\Theta_{_S}|_{{\cal E}}=0 \ , \quad G_{_S}|_{{\cal E}} = g_{_S}=g_{_S}(S,V) \ ,
\ee
so that from the first condition we obtain the first law of thermodynamics and the conditions for thermodynamic
equilibrium in the entropy representation:
\be
\label{eqconds}
d S = \frac{1}{T}d U + \frac{P}{T} dV \ ,\quad
\frac{\partial S}{\partial U} = \frac{1}{T}\ , \quad \frac{\partial S}{\partial V} = \frac{P}{T}\ .
\ee
Moreover, the metric $g_{_S}$ of the equilibrium space can be calculated in a straightforward manner from the above equations and we obtain
\bea
g_{_S}= & &\left(U\frac{\partial S}{\partial U}\right)^{2k+1}\frac{\partial^2 S}
{\partial U^2} dU^2
+  \left(V\frac{\partial S}{\partial V}\right)^{2k+1}
\frac{\partial^2 S}{\partial V^2} dV^2 \nonumber \\
\label{gdowns}
& +&\left[ \left(U\frac{\partial S}{\partial U}\right)^{2k+1}
+ \left(V\frac{\partial S}{\partial V}\right)^{2k+1} \right] 
\frac{\partial^2 S}{\partial U \partial V} dU dV \  \ .
\eea
Again we see that once the fundamental equation $S=S(U,V)$ is given, the metric of ${\cal E}$ is uniquely determined.  
As in classical thermodynamics, the fundamental equation must satisfy the second law which in the case 
under consideration can be written as \cite{callen} 
\be
\label{second}
\frac{\partial^2 U}{\partial E^a \partial E^b} \geq 0\ , \quad 
\frac{\partial^2 S}{\partial F^a \partial F^b} \leq 0 \ ,
\ee
for the energy and entropy representation, respectively. Here $a, b = 1,2\ ,  E^a = (S,V)$ and $F^a = (U,V)$. 

It is worth mentioning that in the above construction we chose  the extensive variables as coordinates for the equilibrium space 
${\cal E}$ in order to obtain the energy and entropy representation which are the most common approaches used in classical thermodynamics. However, in general it is possible to choose any 2-dimensional subspace of ${\cal T}$ to define the
 equilibrium space ${\cal E}$ which would correspond to a different representation in classical thermodynamics. 
 The Legendre invariance of the phase space ${\cal T}$ has as a
consequence that all possible representations are equivalent and the properties of thermodynamic systems do not depend on the representation.

The above geometric construction can be applied to any thermodynamic system with two degrees of freedom. One only needs to specify the fundamental
equation of the thermodynamic system in order to investigate its geometric 
 properties. However, GTD allows a generalization to include any system with
a finite number of degrees of freedom, say $n$. In this case, the phase space has the dimension $2n+1$ and the subspace of equilibrium states is 
$n-$dimensional. The fundamental Gibbs form and the metrics $G$ and $g$ can be generalized to the $(2n+1)-$dimensional case  in a straightforward
manner \cite{vqs08}.

\section{The equilibrium space of the ideal gas}
\label{sec:ig}

In the specific case of an ideal gas, the fundamental equation in the entropy representation 
can be expressed as \cite{callen}
\be
\label{feig}
S(U,V) = S_0 + N k_{_B} c_{_V}\ln\left(\frac{U}{ U_0}\right)  + N k_{_B} 
\ln\left(\frac{V}{ V_0}\right) \ ,
\ee
where $c_{_V}$ is the dimensionless 
heat capacity at constant volume, $k_{_B}$ is Boltzmann's constant, 
$N$ is the constant number of particles and $S_0$, $U_0$ 
and $V_0$ are constants.

The intensive thermodynamic variables for the ideal gas can be calculated by using the conditions
of thermodynamic equilibrium (\ref{eqconds}). We obtain
\be
\label{steq}
\frac{1}{T} = \frac{N k_{_B} c_{_V}}{U}\ ,\quad
\frac{P}{T} = \frac{N k_{_B}}{V}\ .
\ee

Furthermore, from Eq.(\ref{gdowns}) we obtain the simple metric 
\be
\label{gdownig}
g_{_S} = -(N k_{_B})^{2k+2} \left[ c_{_V}^{2k+2}\frac{dU^2}{U^2} +  \frac{dV^2}{V^2}\right]\ .
\ee
It is straightforward to show that the curvature of this metric vanishes. 
In GTD, we interpret the absence of curvature  as a manifestation of the absence of thermodynamic
interaction. One of the goals of GTD is to interpret the curvature of the space of equilibrium states as a measure of thermodynamic interaction.
This goal has been reached here in the case of the ideal gas. 
Moreover, one can show \cite{quev07,qr11} that in the case of 
the van der Waals gas the curvature tensor of the metric (\ref{gdowns}) does not vanish in accordance with the fact that the van der Waals gas corresponds
to a system with non-vanishing thermodynamic interaction. In addition, the locations of the thermodynamic curvature singularities turn out to coincide 
with the points where phase transitions occur \cite{qr11}. We interpret this result as a further indication that thermodynamic curvature
can be used to measure the thermodynamic interaction. 

Since the curvature of the metric (\ref{gdownig}) vanishes, there must exist coordinates in which the metric takes the simple Euclidean
form. In fact, the Cartesian-like coordinates   
\be
\xi = \xi_{int} + (N k_{_B}c_{_V})^{k+1}   \ln U \ , \quad \eta = \eta_{int} + (N k_{_B})^{k+1} \ln V \ 
\ee
lead to the metric 
\be
\label{flat}
g_{_S} = d\xi^2 + d\eta^2 \ , \quad 
\ee
where for simplicity we have chosen a positive definite signature. Moreover, the additive constants of integration  and  $\xi_{int}$ and $\eta_{int}$ can always be chosen such that $\xi \geq 0$ and $\eta \geq 0$. This means that the equilibrium space of the ideal case can be represented as the positive definite quadrant of the Cartesian plane. 

\section{Thermodynamic geodesics}
\label{sec:geoig}

\begin{figure}
\begin{center}
\includegraphics[width=7cm]{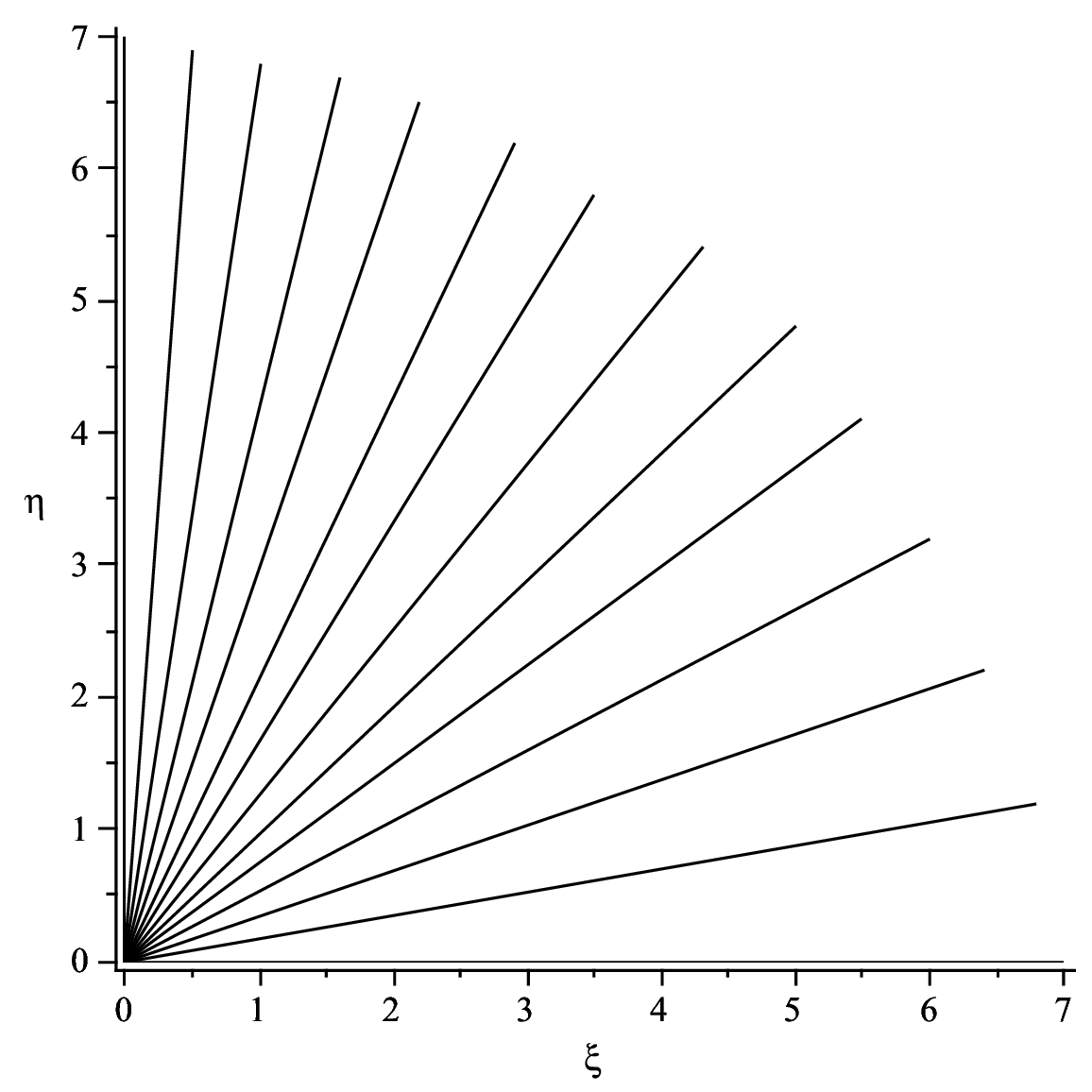}
\end{center}
\caption{Geodesics in the space of equilibrium states of the ideal gas in logarithmic 
coordinates
$\xi \propto \ln U $ and $\eta \propto \ln V$. The initial state is located on the origin $(0,0)$.  
In general, two arbitrary equilibrium states can always be connected by means of a geodesic.}
\label{fig:geo1}
\end{figure}

Let us define in the equilibrium space ${\cal E}$ the thermodynamic length as  
$L = \int \sqrt{g_{_S}} dE$, where $dE$ is the volume element in  ${\cal E}$ \cite{quev07,vqs08}. 
Then, the condition that the thermodynamic length be extremal, $\delta L =0$, leads to the geodesic equations for the
coordinates of the space ${\cal E}$. In the case of the flat metric (\ref{flat}), 
the geodesic equations become $\ddot \xi =0$ and $\ddot \eta = 0$, where the dot denotes differentiation
with respect to an affine parameter $\tau$. 
The solutions are found to be  $\xi = \xi_1 \tau+ \xi_0$
and   $\eta = \eta_1 \tau + \eta_0$, where $\xi_0,\ \xi_1, \ \eta_0$ and $\eta_1$
are constants,  i. e., they represent straight lines.
For instance, 
consider all geodesics with initial state $\xi_i=0$ and $\eta_i=0$. Then,  
on the $\xi\eta-$plane the geodesics 
must be contained within the quadrant  determined by  $\xi\geq 0 $ and $\eta \geq 0$, due
to our choice of integration constants for $\xi$ and $\eta$. Consequently, the geodesics of the ideal gas
can be depicted by using the equation $\xi=c_1\eta + c_0$, 
with constants $c_0$ and $c_1$. For any arbitrary initial state,  
there is always a straight line that connects that state with any arbitrary point on the $\xi\eta-$plane.
This means that the entire space of equilibrium states can be covered by geodesics. 
This behavior is schematically 
illustrated in Fig. \ref{fig:geo1}.

However, not all the solutions of the geodesic equations must be physically realistic.  
Indeed, there could be straight lines 
connecting equilibrium states that are not compatible with the laws of thermodynamics.
In particular, one would expect that the second law of thermodynamics imposes strong requirements on the solutions.  
In classical thermodynamics, two equilibrium states are related to each other only if they 
can be connected by means of a quasi-static process. Then, a geodesic that connects two physically
meaningful equilibrium states can be interpreted as representing a quasi-static process. 
Since a geodesic curve is a dense succession of points, we conclude
that a quasi-static process can be seen as a dense succession of equilibrium states, a statement
which coincides with the definition of quasi-static processes in equilibrium thermodynamics  \cite{callen}.
Accordingly, we  define a thermodynamic geodesic as a geodesic along which the laws of thermodynamics are satisfied, i.e., a 
geodesic that represents a quasi-static process. This implies that a thermodynamic geodesic must have a definite direction 
associated with the direction in which the entropy increases, in agreement with the second law of classical thermodynamics. 
Moreover, since the affine parameter is defined up to a linear transformation, 
it should be  possible to choose it  in such a way that it increases as the 
entropy of a quasi-static process increases. This opens the possibility of interpreting the
affine parameter as a ``time" parameter with a specific direction which coincides with the
direction of entropy increase. We will explore in detail this possibility in Sec. \ref{sec:cone}.

\begin{figure}
\begin{center}
\includegraphics[width=7cm]{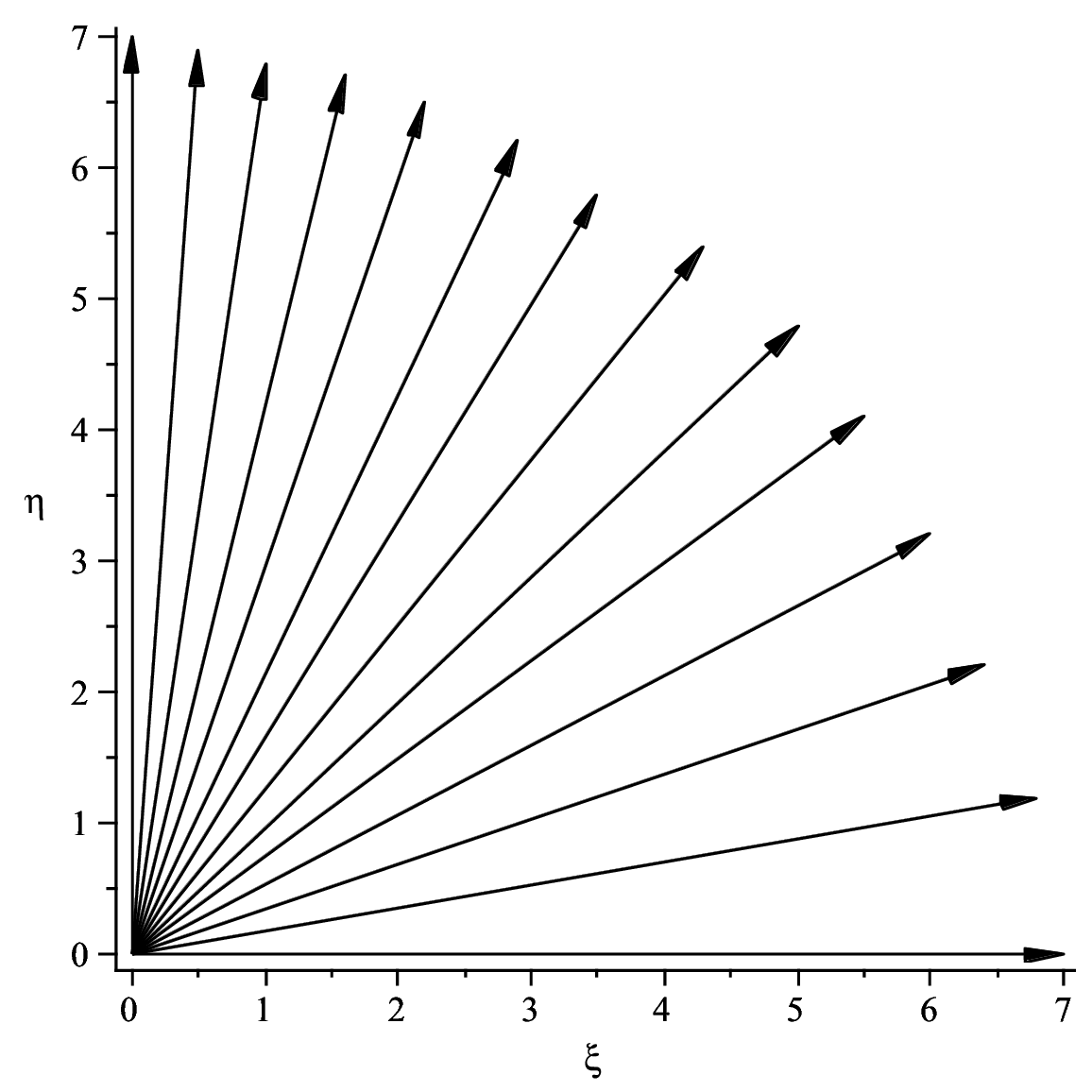}
\end{center}
\caption{Geodesics that satisfy the second law of thermodynamics. The initial equilibrium state
is located on the origin of coordinates. There is only one geodesic which connects the origin with any 
other equilibrium state. The arrows show the direction in which entropy increases, suggesting that they 
could be interpreted as the ``arrows of time"'.  
}
\label{fig:geo2}
\end{figure}

In the special case of the ideal gas, the fundamental equation (\ref{feig}) 
in coordinates $\xi$ and $\eta$ represents a straight line (with a new additive constant which we denote as before as $S_0$ for simplicity)
\be
\label{feqigs}
S = S_0  + (N k_{_B} )^{-k} \left( c_{_V}^{-k}  \xi + \eta \right) \ ,
\ee
so that the entropy increases as $\xi$ and $\eta$ increase. 
Consequently, any straight 
line pointing outwards of the initial zero point and contained inside 
the allowed positive quadrant connect states with increasing entropy. 
This behavior is schematically 
illustrated in Fig. \ref{fig:geo2} where the arrows indicate the direction 
in which a quasi-static process can take place, i.e., in which the entropy 
increases. A quasi-static process connecting
states in the opposite direction is not allowed by the second law of thermodynamics. 
Consequently, the affine parameter $\tau$ represents a 
time parameter and the direction on each geodesic indicates 
the ``arrow of time". 

Since the above description is based upon a rigorous analysis of the
geodesic equations and their solutions in the equilibrium space of an ideal gas, 
we conclude that the above result can be interpreted as a rigorous geometric proof of the
intuitive and well-known idea that the direction of time coincides with the
direction in which entropy increases. It must 
be mentioned that our interpretation of ``time" is pure classical in the sense
that it corresponds to an affine parameter $\tau$ along a macroscopic geodesic, and
cannot be extrapolated to the microscopic level where time must be 
interpreted in a different manner and possibly quantum effects must be considered  \cite{huang,landau}.

If the initial state is not at the origin of the $\xi\eta-$plane,
the second law permits the existence of geodesics for which one of the coordinates,
say $\eta$, decreases as long as the other coordinate $\xi$ increases in such a
way that the entropy increases or remains constant. 
In fact, the region in the $\xi\eta-$plane available from a given initial equilibrium state is 
situated within two extreme geodesics which span a maximum angle that can be determined in the following 
way. Let the initial state be at the point $(\xi_i,\eta_i)$. According to the second law of thermodynamics and 
Eq.(\ref{feqigs}),
the state characterized by the coordinate values
$(\xi_f, \eta_f)$ can be reached by a geodesic with origin at $(\xi_i,\eta_i)$  
if the condition 
\be
\label{entcond}
c_{_V}\Delta \xi  +  \Delta\eta\geq 0 \ ,\quad \Delta \xi = \xi_f -  \xi_i\ ,\quad \Delta \eta = \eta_f -  \eta_i\ ,
\ee
is satisfied. Consider geodesics for which $\Delta\xi < 0$. Hence, only those geodesics are allowed for which 
$\Delta \eta \geq c_{_V}|\Delta\xi|$. The equal sign determines the extreme geodesic with constant entropy (adiabatic geodesic) which intersects the
$\eta-$axis at the point $\xi_f=0$ and $\eta_f = \eta_i + c_{_V}|\Delta\xi| = \eta_i + c_{_V} \xi_i$. This geodesic intersects
the  $\eta-$axis at an angle $\alpha$ such that $\tan\alpha = 1/c_{_V}$.  
Consider now geodesics with $\Delta\eta < 0$. The allowed
geodesics must satisfy $\Delta\xi \geq |\Delta \eta|/c_{_V}$ 
and the extreme adiabatic geodesic intersects the $\xi-$axis at 
the point with coordinates $\eta_f = 0$ and $\xi_f = \xi_i + |\Delta\eta|/c_{_V} 
= \xi_i + \eta_i/c_{_V}$. The adiabatic geodesic 
intersects the $\xi-$axis at an angle $\alpha^\prime$ with $\tan\alpha^\prime = c_{_V}$. 
Since the intersection angles are 
complementary, $\tan\alpha^\prime = 1/\tan\alpha$, we conclude that the angle spanned by  the two adiabatic geodesics (one  
with $\Delta\xi <0$ and the second one with  $\Delta\eta <0$) is $\pi/2$.

\begin{figure}
\begin{center}
\includegraphics[width=7cm]{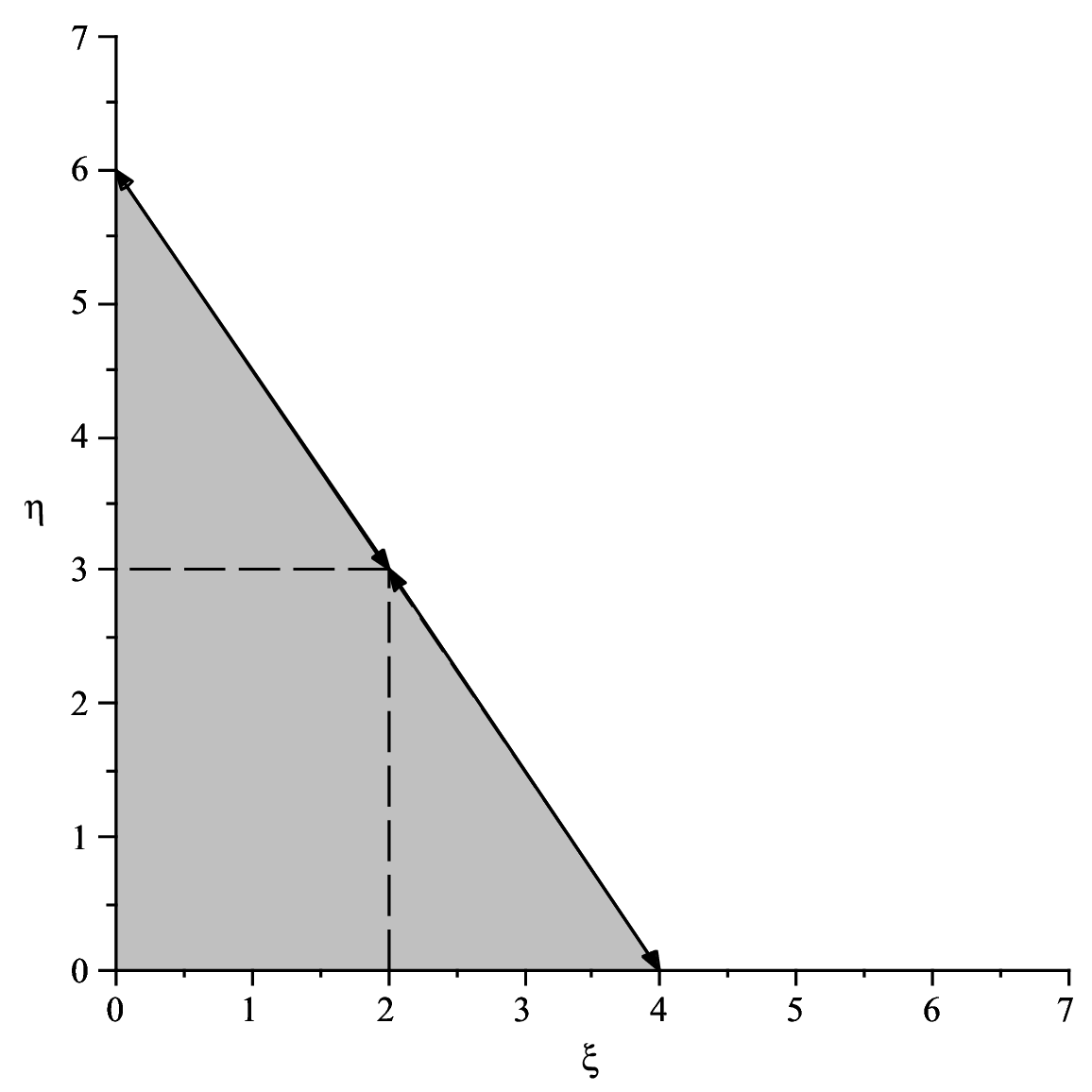}
\end{center}
\caption [] {Adiabatic geodesics with initial state at $\eta_i=3$, $\xi_i=2$. They represent reversible processes so 
that the ``arrow of time" can point in both directions. States in the shadow region are connected to the initial state
by geodesics with a negative change in the entropy.
}
\label{fig:geo3}
\end{figure}

An alternative derivation of the above geometric construction of adiabatic geodesics consists in considering 
the corresponding equation in the form $c_{_V}(\xi_f -  \xi_i)  + \eta_f -  \eta_i = 0$, which can be rewritten 
as
\be
\frac{\xi_f}{\xi_i + \eta_i / c_{_V}} + \frac{\eta_f}{\eta_i + \xi_i c_{_V}} = 1 \ ,
\ee
and is immediately recognized as the equation of a straight line. This line in the equilibrium space can be occupied only 
by states belonging to an adiabatic process. Moreover, since the entropy remains constant along this straight line, the 
``arrow of time" can point in both directions. This is illustrated in Fig. \ref{fig:geo3}.
From the last equation it is then easy to obtain the general relationships
\be
\tan\alpha = \frac{\xi_i + \eta_i/c_{_V}}{\eta_i + \xi_i c_{_V}}\ ,\quad
\tan\alpha^\prime = \frac{\eta_i + \xi_i c_{_V}}{\xi_i + \eta_i/c_{_V}}\ .
\ee
These formulas are valid for any values of the initial state, except the one situated on the origin of coordinates. 
In fact, for the initial 
states $(0,\eta_i)$ and $(\xi_i, 0)$, we recover the values of the intersection angles described above.

If the initial state 
is on the origin, the entropy condition (\ref{entcond}) is always satisfied since any arbitrary straight line that 
starts at the origin is characterized 
by $\Delta\xi \geq 0$ and $\Delta\eta \geq 0$ so that the allowed geodesics could occupy the entire positive quadrant as illustrated
in Fig. \ref{fig:geo2}. 
However, this result changes drastically if we take into account the third law of thermodynamics which postulates 
the impossibility of reaching absolute zero of temperature or, equivalently, the minimum value of the entropy \cite{callen}. 
For the ideal gas with
fundamental equation (\ref{feqigs}) the minimum value for the entropy is $S_0$ and corresponds to $\xi=0$ and $\eta=0$. Consequently,
the origin of coordinates must be ``removed" from the space of equilibrium states. 

We conclude that in the case of the ideal gas, a thermodynamic geodesic can be represented
as a straight line that never crosses
the origin of coordinates and possesses a definite direction which  coincides with the direction of entropy increase. 

\begin{figure}
\begin{center}
\includegraphics[width=7cm]{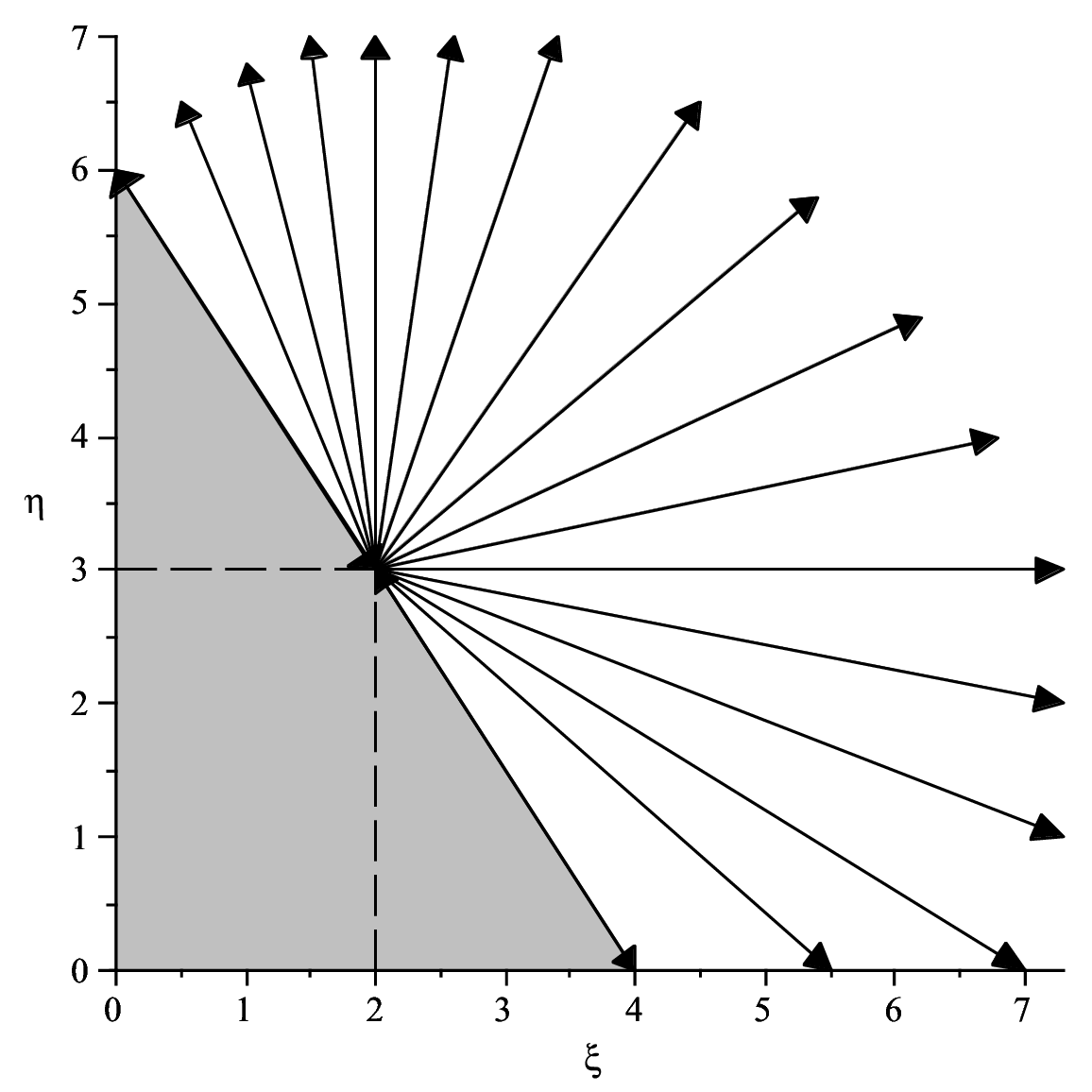}
\end{center}
\caption [] {General structure of the thermodynamic geodesics in the space of equilibrium states of an ideal gas. 
 Here we choose a monoatomic gas so that 
$c_{_V} = 3/2$. Consequently, $\alpha\approx 33.3^o$ and $\alpha^\prime \approx 56.7^o$.  
The shadow region contains all the states that due to the second law of thermodynamics  
cannot be reached by thermodynamic geodesics with the fixed initial state. Adiabatic geodesics
determine the boundary of the connectivity region where several thermodynamic geodesics are depicted.
}
\label{fig:geo4}
\end{figure}

We see that the laws of thermodynamics imply that the geometric structure of the equilibrium space is as illustrated 
in Fig. \ref{fig:geo4}. For any given initial equilibrium state $(\xi_i,\eta_i)$, there exist two different regions. 
The first one is determined by all the states than can be reached from the initial state by means of quasi-static processes, i. e.,
by thermodynamic geodesics. This could be called the {\it region of connectivity} of the  initial state. 
If we identify $\tau$ as a time parameter, the connectivity region acquires the characteristics of a
causally-connected region, resembling concepts of relativistic
physics. The second region is composed of all the equilibrium states that cannot be reached from the initial state by thermodynamic
geodesics. We call it the {\it region of non-connectivity}. Again, it could be also identified with the non-causally 
connected region of spacetime in 
relativistic physics. The boundary between the connectivity and non-connectivity regions is occupied 
by adiabatic thermodynamic geodesics and this is the only place in the equilibrium space where reversible thermodynamic processes can 
occur. 

\section{The adiabatic cone}
\label{sec:cone}

As mentioned in the previous section, it is possible to use the entropy as an affine parameter. Indeed, introducing the geodesic solutions 
$\xi=\xi_0 + \xi_1 \tau$ and  $\eta=\eta_0 + \eta_1 \tau$ in Eq.(\ref{feqigs}), we obtain $S=\tilde S _0 + \tilde S_1 \tau$ so that the affine 
parameter can be represented as $\tau = \tau_0 + \tau_1 S$, where $\tau_0$ and $\tau_1$ are constants. This implies that in terms of $S$ the logarithmic coordinates
can be expressed as
\be
\xi=\tilde \xi_0 + \tilde \xi_1 S\ , \quad \eta=\tilde \eta_0 + \tilde \eta_1 S \ ,
\ee
where
\be
\tilde\xi_0 = \frac{\xi_0\eta_1 - \xi_1 \eta_0 -  S_0 (Nk_{_B})^k \xi_1}{c_{_V}^{-k} \xi_1 + \eta_1 }\ , \quad 
\tilde\xi_1 = \frac{(N k_{_B})^k \xi_1}{c_{_V}^{-k} \xi_1 + \eta_1}\ ,
\ee
\be
\tilde\eta_0 = \frac{ c_{_V}^{-k} (\xi_1\eta_0 - \eta_1\xi_0) - S_0 (N k_{_B})^k \eta_1  }{ c_{_V}^{-k} \xi_1 + \eta_1}\ , \quad
\tilde\eta_1 = \frac{(N k_{_B})^k \eta_1}{c_{_V}^{-k} \xi_1 + \eta_1}\ .
\ee
It is easy to see that the new constants $\tilde \xi_0$, $\tilde \xi_1$, $\tilde \eta_0$, and $\tilde \eta_1$ can always be taken as positive definite by choosing 
appropriately the original constants $\xi_0$, $\xi_1$, etc. 

\begin{figure}
\begin{center}
\includegraphics[width=10cm]{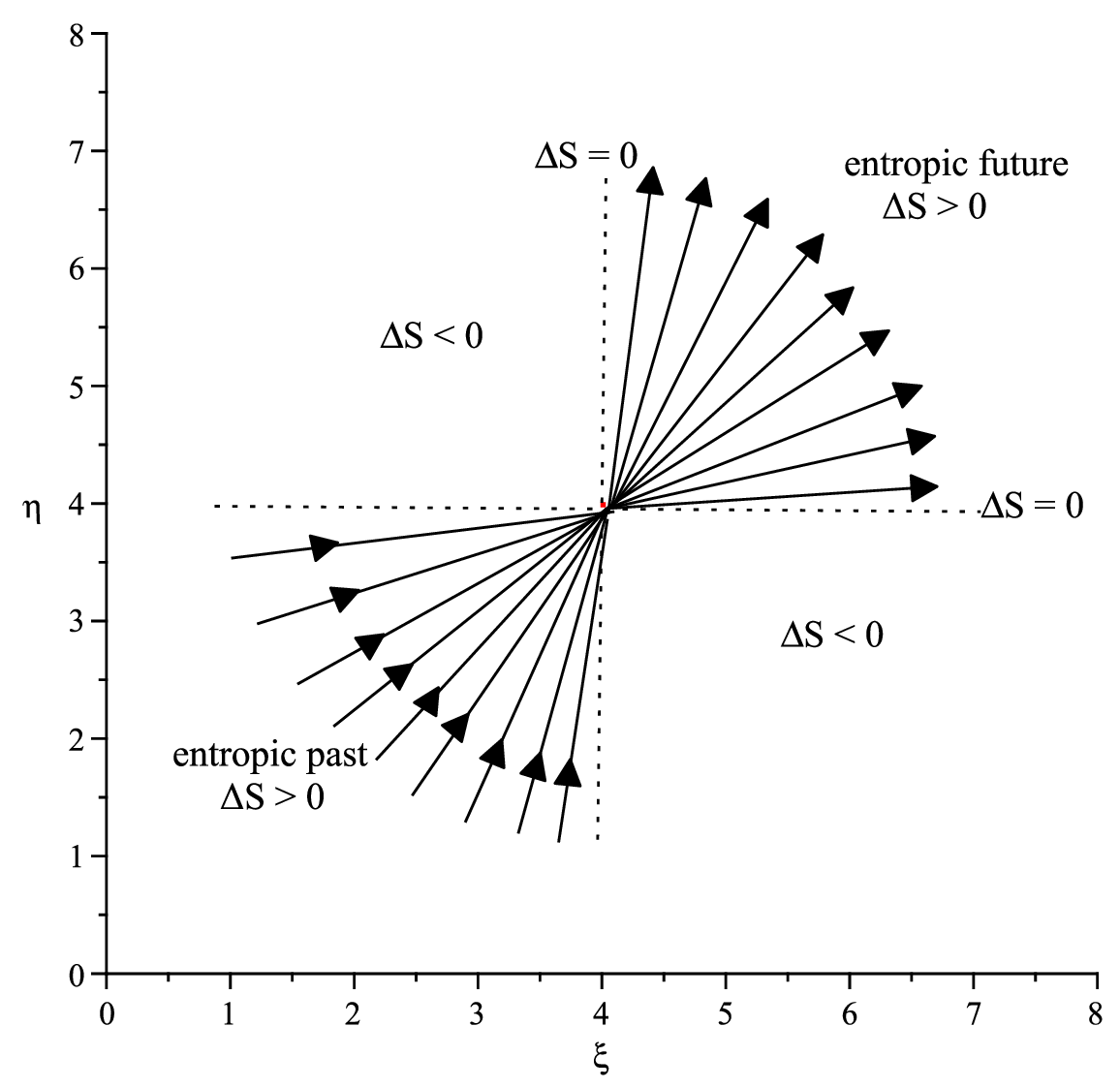}
\end{center}
\caption [] {The adiabatic cone of a particular equilibrium state. No thermodynamic geodesic can exist on the boundaries $\Delta S=0$. The arrows indicate the direction in which the entropy increases, i.e., in which quasi-static processes occur. The direction of the thermodynamic geodesics can also be interpreted as the 
``arrow of time".   }
\label{fig:geo6}
\end{figure}

In this representation, an initial equilibrium state with coordinates $(\xi_i,\eta_i)$ is connected to a final state  $(\xi_f,\eta_f)$ by a thermodynamic geodesic
if the conditions 
\be
\Delta \xi = \xi_f -\xi_i =\tilde \xi_1 \Delta S \geq 0\ , \quad
\Delta \eta = \eta_f -\eta_i =\tilde \eta_1 \Delta S \geq 0\ , 
\ee
are satisfied. Since the second law demands that $\Delta S \geq 0$, we conclude that all the thermodynamic geodesics must satisfy the conditions $\Delta\xi\geq 0$ and
$\Delta\eta\geq 0$, simultaneously. Consequently, all the thermodynamic geodesics that initiate at a particular equilibrium state must be contained within the region  defined by $\Delta\xi =0$ and $\Delta\eta =0$.  On the other hand, all the thermodynamic geodesics that end at a particular equilibrium state, say $(\xi_i,\eta_i)$ 
must obey the same conditions. Figure \ref{fig:geo6} illustrates this behavior. We conclude that all the ``incoming" and ``outgoing" thermodynamic geodesics at a given point must be contained within a ``cone" that we will call {\it adiabatic cone}. 

Notice that the boundaries of the adiabatic cone are determined by adiabatic processes ($\Delta S =0)$ which in this representation must obey the condition $\Delta\xi=\Delta\eta=0$, i. e., they are confined to one point in the equilibrium space. We designate as entropic future of an equilibrium state the region of the equilibrium space occupied by the thermodynamic geodesics that start at that state. The entropic past is defined in a similar way with all the thermodynamic geodesics that end at that particular equilibrium state.

\section{Discussions and conclusions}
\label{sec:con}

In this work, we presented a detailed analysis of the geometry of the ideal gas equilibrium space. The main 
property of our analysis is its invariance with respect to Legendre transformations, i. e.,
it is independent of the choice of thermodynamic potential, a property which is essential
in classical thermodynamics. The geometry of the ideal gas turns out to be flat, in 
accordance with our intuitive expectation that the absence of thermodynamic interaction
would imply absence of curvature. This is an indication that thermodynamic curvature can 
be used as a measure of thermodynamic interaction. 

Our analysis of the geodesics in the space of equilibrium space shows that they can be represented
as straight lines when logarithmic thermodynamic variables are used. We introduce the concept 
of thermodynamic geodesics as those solutions of the geodesic equations which satisfy the laws
of thermodynamics.  Then, the equilibrium space can be represented as a 
Cartesian-like plane where thermodynamic geodesics correspond to quasi-static processes. 
The third law of thermodynamics implies that the origin of coordinates must be removed from 
the space of equilibrium states. This opens the possibility of interpreting the third law 
as a topological property of the equilibrium space. 

For any given initial state, the equilibrium space can be split into two different regions.
The connectivity region is occupied by all states which are connected to the initial state by 
means of thermodynamic geodesics. On the contrary, the region of non-connectivity corresponds
to those equilibrium states than cannot be reached from the initial state by using only 
thermodynamic geodesics. In the boundary between these two regions there exist adiabatic 
thermodynamic geodesics. It can be shown that this is the only place where adiabatic geodesics
can exist so that the boundary determines the only region in the equilibrium space
where reversible thermodynamic processes can take place. Moreover, the  concept of 
``arrow of time" becomes well-defined in terms of thermodynamic geodesics. 

If we use the entropy as the affine parameter along the thermodynamic geodesics, we have shown that the connectivity region of a
given equilibrium state can be represented as an adiabatic cone inside which the thermodynamic geodesics have a definite direction, 
which coincides with the direction of entropy increase, and can be interpreted as the ``arrow of time". 
Since our results are based upon
a rigorous analysis of the geodesic equations and their solutions, we conclude that 
we have provided a geometric rigorous proof of the intuitive and well-known idea that the 
``arrow of time" coincides with the direction in which entropy increases. Of course, we have
shown this explicitly in this work only for the case of an ideal gas. 
Whether this conclusion 
holds also in the case of more realistic thermodynamic systems remains a task for future 
investigations. However, from a geometric point of view this seems to be true in general. In fact, since any differential manifold can be 
interpreted as locally flat, all the results obtained here for the ideal gas should be valid locally for any thermodynamic system.

In this work, we used the entropy representation. For the sake of completeness, we performed 
the same analysis using the energy representation. The main difference in the analysis follows
from the fact that the fundamental equation (\ref{feig}) in the energy representation, 
$U(S,V) = U_0 \exp(S/N k_{_B}c_{_V})/V^{c_{_V}}$, leads to a non-diagonal term in the metric of the 
space  of equilibrium states. This complicates the analysis of the geodesic equations. Nevertheless, 
the corresponding metrics with all possible thermodynamic potentials were shown to describe 
a flat space so that  in all the cases it is possible to introduce Cartesian-like coordinates 
and the investigation of the geodesic equations leads to results equivalent to those obtained in 
the entropy representation. This result corroborates one of the most important properties of the
formalism of GTD, i. e., the choice of thermodynamic potential or representation does not 
affect the results. This is also explicitly shown in the Appendix. 

We found that the space of equilibrium states of the ideal gas possesses a very rich geometric 
structure which resembles the structure of spacetime in relativistic physics. Since this 
structure is the result of applying the laws of thermodynamics in a geometric context, we  
expect similar structures in the case of more general thermodynamic systems. For 
instance, we analyzed the van der Waals gas  which is generally accepted as 
describing the thermodynamics of realistic gases. In this case 
the metric for the equilibrium space can be obtained immediately 
from Eq.(\ref{gdowns}) by introducing the corresponding fundamental equation $S=S(U,V)$. 
The resulting metric, however, is no longer flat. This is due to the fact that the van der 
Waals gas possesses a certain degree of thermodynamic interaction which, according to the 
formalism of GTD, generates a non-vanishing thermodynamic curvature. It turns out that
the corresponding curvature scalar possesses singular points which coincide with the 
points of phase transitions of the van der Waals gas. 
Moreover, as a result of the non-vanishing curvature, the geodesic
equations are much more complicated and there are no Cartesian-like coordinates in which the 
geodesics could be represented as straight lines. The geodesic equations must be analyzed by using numerical methods \cite{qr11}. 

In this work, we limit ourselves to systems with only two thermodynamic degrees of freedom. 
We can increase the number of degrees of freedom, maintaining the flatness of the equilibrium
space, by considering, for instance, a multicomponent ideal gas, a paramagnetic ideal gas, etc.
In fact, it can be shown explicitly that the formalism of GTD, as described in \ref{sec:gtdbh},
can be generalized to include systems with a larger number of thermodynamic variables. 
As a result, one obtains a metric, similar to (\ref{gdowninv1}), for higher dimensional manifolds  
from which one can show that the equilibrium spaces for the above mentioned generalizations of the ideal gas
are flat. In all these cases, the flatness of the equilibrium space implies that there exist Cartesian-like
coordinates in which the thermodynamic geodesics can be represented again as straight lines 
in higher dimensional spaces. 
The splitting of 
the equilibrium space in a region of connectivity separated by adiabatic geodesics from 
the non-connectivity region is similar; however, those regions are now represented by spaces
with dimensions higher than 2.
We conclude that the analysis of the corresponding thermodynamic geodesics will require 
to consider the topological and geometric properties of equilibrium spaces in higher dimensions.

The results presented in this work suggest that the flat equilibrium space of the ideal gas is to GTD what the Minkowski spacetime 
is to general relativity. We found an analogy between the thermodynamic structure of the equilibrium space as determined by thermodynamic geodesics  and the causality structure of the Minkowski spacetime as determined by test particles. This is a new aspect of the inter-relationship between thermodynamics and relativistic physics.

\section*{Acknowledgements} 
We would like to thank the GTD-UNAM group for helpful comments and discussions. One of us (HQ) thanks 
F. Schaller for stimulating discussions. 
This work was supported DGAPA-UNAM, Grant No. 113514, and Conacyt-Mexico, Grant No. 166361. 

\appendix
\section{The entropy representation}
\label{sec:app1}
As mentioned in Section \ref{sec:ig}, in the entropy representation the coordinates
of the phase space ${\cal T}$ 
are $ (S, U, V, \beta, \vartheta)$, with $\beta =1/T$ and 
$\vartheta = P/T$ so that the auxiliary metric in ${\cal T}$ can be written 
as
\be
\label{gups1}
G_{_S} = \left(dS -\beta d U - \vartheta dV\right)^2
+  \left[\left(U \beta\right)^{2k+1} dU d\beta  
+\left(V\vartheta\right) ^{2k+1} dV d\vartheta \right] \ .
\ee
As before, the structure of this auxiliary metric is such that 
any extensive variable is multiplied by its corresponding intensive variable. As
a result the above metric is invariant with respect to the following
Legendre transformations,  
 $ (S, U, V, \beta, \vartheta) \longrightarrow 
(\tilde S, \tilde U, \tilde V, \tilde \beta, \tilde \vartheta)$: 
\be
\tilde S_1 = S - U \beta \ ,\quad 
U = -\tilde \beta \ ,\ \beta = \tilde U \ , 
V = \tilde V \ ,\quad  \vartheta = \tilde\vartheta \ ,
\ee
\be
\tilde S_2 = S - V \vartheta \ , \quad
U = \tilde U \ , \beta = \tilde \beta \ , \quad
V = -\tilde \vartheta\ , \quad \vartheta = \tilde V \ ,
\ee
\be
\tilde S_3 =  S - U \beta - V \vartheta \ , \quad
U = -\tilde \beta \ ,\ \beta = \tilde U \ , 
V = -\tilde \vartheta\ , \quad \vartheta = \tilde V \ .
\ee
The thermodynamic potentials $\tilde S_1$, $\tilde S_2$, and $\tilde S_3$ are known 
as Massieu functions \cite{callen}. For a given fundamental equation $S=S(U,V)$ 
they represent the same subspaces of ${\cal T}$ in different coordinates. 
In the case of the ideal gas, the Massieu functions can be derived explicitly 
by using the state equations (\ref{steq}) in the form $U\beta = c_{_V} N k_{_B}$ and 
$V\vartheta = N k_{_B}$ and the fundamental equation (\ref{feig}). We get (dropping 
the tildes) 
\be
S_1(\beta,V)   = S_{01} - c_{_V}N k_{_B}\ln \beta + N k_{_B} \ln V \ ,
\ee
\be
S_2(U,\vartheta)   = S_{02} + c_{_V}N k_{_B}\ln U -  N k_{_B} \ln \vartheta \ ,
\ee
\be
S_3(\beta,\vartheta)   = S_{03}  - c_{_V}N k_{_B}\ln \beta -  N k_{_B} \ln \vartheta \ ,
\ee
where $S_{01}$, $S_{02}$, and $S_{03}$ are constants. Then, the metric of the 
subspace of equilibrium states ${\cal E}$ in each case corresponds to
\be
g_1 = \frac{d\beta^2}{\beta^2} + \frac{dV^2}{V^2}\ ,
\ee
\be
g_2= \frac{d U ^2}{U ^2} + \frac{d\vartheta^2}{\vartheta^2}\ ,
\ee
\be
g_3= \frac{d\beta^2}{\beta^2} + \frac{d\vartheta^2}{\vartheta^2}\ ,
\ee
where for simplicity we omit the constants and choose a positive definite signature.  Clearly, 
all these metrics represent the same flat space of equilibrium states for the ideal
gas. The geodesic equations can be solved and 
we obtain $Z^I = Z^I_0 \exp(\tau/\tau_{_I})$, where 
$Z^I = (U,V,\beta,\vartheta)$ and $Z^I_0$ and $\tau_{_I}$ are constants. An analogous representation for the 
geodesics can be obtained from the analysis performed in Sec. \ref{sec:geoig}.  
This is a concrete example of the invariance of the results obtained by using the formalism
of GTD.


\end{document}